\documentclass[12pt]{iopart}
\usepackage{epsfig}
\begin{document}

\title[Stabilization of Burn Conditions ...]{ 
Stabilization of Burn Conditions in an ITER FEAT like
 Tokamak with Uncertainties in the Helium Ash
Confinement Time }

\author{Javier E. Vitela  
\footnote[3]{E-mail address: vitela@nuclecu.unam.mx}
}

\address{Instituto de Ciencias Nucleares,
 Universidad Nacional Aut\'onoma de M\'exico \\ 
 04510 M\'exico D.F.}

%\maketitle
\begin{abstract}
In this work we demostrate  using a two-temperature volume averaged 0D
  model that robust stabilization, with regard the helium ash confinement
time, of the burn conditions of a tokamak
 reactor with the  ITER FEAT design parameters 
  can be achieved using Radial
 Basis Neural Networks (RBNN).   
   Alpha particle thermalization time delay is 
 taken into account in this model.  
The control actions implemented by means of a RBNN, include the modulation
 of the DT refueling rate,  a neutral He-4 injection beam and auxiliary
 heating powers to ions and electrons; all of them constrained to lie
 within allowable range values.   Here we assume that the tokamak follows
the IPB98(y,2) scaling for the energy confinement time, while helium
 ash confinement time is assumed to be independently estimated on-line.
 The DT and helium ash particle confinement times are assumed to keep 
 a constant
 relationship at all times.    An on-line noisy estimation
 of the helium
ash confinement time due to  measurements is  simulated by
 corrupting it with pseudo Gaussian noise.
\end{abstract}

%Uncomment for PACS numbers title message
%\pacs{00.00, 20.00, 42.10}

% Uncomment for Submitted to journal title message
%\submitto{\JPA}

% Comment out if separate title page not required
%\maketitle

\section{Introduction}
In a burning regime a reactor plasma must be heated mainly by the
energetic particles produced by fusion. In DT fueled reactors in
particular the $\alpha$-particles produced will deposit, during
slowing down, most of 
their energy to the plasma electrons. The highly energetic
alpha particles are expected to destabilize MHD modes known are
Alfven eigenmodes.  The strong  
 nonlinear coupling among
the energy deposition profile of the alpha particles, the
new MHD instabilities,
the bootstrap current
 and the plasma boundary, will make
  transport properties  significatively different from those
 observed in current tokamak experiments.[1] 
Hence, active control of particle densities and plasma temperature
will be essential in order to regulate the power density and to suppress
fluctuations in plasma parameters due to turbulence and/or changes in
confinement modes.

 Here we report the results of a burn control study
of an ITER FEAT-like tokamak by means of radial basis
artificial neural networks with Gaussian nodes in the hidden layer and
sigmoidals in the output layer using
a two-temperature volume-averaged 0-D model,[2] assuming
the particle density is homogeneous throughout the plasma core, with
 electrons and ions having the same radial profile  but
 different peak temperatures.
 In contrast with previous works[3]  alpha particle
 thermalization time delay is 
 taken into account in this model.  
It is  assumed  that the energy confinement time
of the reactor follow the IPB98(y,2) scaling and that
 the helium ash particles confinement time $\tau_\alpha$,
 is  independently  estimated
"on-line".  Their
current estimated value contains noise due to turbulence and/or intrinsic
 measurement uncertainties, and is
 fed, together with the electron and ions temperatures, the electron density
 and
the helium ash fraction, into the RBNN controller.
  The control actions are implemented through the concurrent modulation of
the refueling rate $S_f$,  the
neutral He-4 injection rate $S_\alpha$, and
the auxiliary heating power density deposited to the ions and the
electrons, $P_{\hbox{\footnotesize aux},i}$ and
 $P_{\hbox{\footnotesize aux},e}$, respectively, which can take values
only within appropiate minima and maxima.

\section{Model}
The fusion reactor  model  considered here
 describes the time evolution of a quasineutral plasma composed
of electrons, 50:50 D-T fuel, helium ash,  a small amount of
 Be and Ar impurities,  whose densities are
$n_e$, $n_{DT}$, $n_\alpha$,
$n_{\hbox{\footnotesize Be}}$ and $n_{\hbox{\footnotesize Ar}}$,
respectively.
The total thermal energy is determined assuming
Maxwellian distribution of the particles:  the electrons with a temperature
profile $T_e(r,t)$, and all the ions 
with the same radial
 profile  $T_i(r,t)$.
The plasma heating takes place mainly by the thermalization of the
alpha particles produced by the fusion
  reactions together with an external RF electron and ion heating,
 with a small contribution of joule heating.
Bremsstrahlung is the only radiation
loss mechanism considered.
We  assume that both
the density  and the effective charge  of the impurities particles
remain constant at all times. The simple model used here,
 before volume average is taken, is represented by the  following coupled
set of equations
                                                                                
%------------------------------------
\normalsize
\begin{eqnarray}
\hspace{-15truemm}{\partial \over \partial t} n_{DT}  =  \   S_f \  -  \
{1 \over 2} n_{DT}^2
 < \sigma v > \  -  \ \nabla \cdot \vec \Gamma_{DT}  &&  ,  \\
 \vspace{+12.truemm}
\hspace{-13truemm} {\partial \over \partial t} n_\alpha \    =   \   S_\alpha \  +  \
{1 \over 4} f_{\hbox{\footnotesize frac}} \ 
 \int_0^\infty dt^\prime 
 \xi_{\hbox{\footnotesize th}}(t^\prime) 
 n_{DT}^2(t-t^\prime)  &&
\langle \sigma v(t-t^\prime) \rangle  \ - \  \nabla \cdot \vec \Gamma_\alpha
 ~~ ,
\end{eqnarray}
\begin{eqnarray}
 \hspace{-15.7truemm} {\partial \over \partial t} \biggl[ {3 \over 2}   n_e T_e\biggr]    =
 &&  P_{\hbox{\footnotesize aux},e}   +  {1 \over 4} \
 f_{\hbox{\footnotesize frac}}
 \ f_e \ Q_\alpha
 \int_0^\infty dt^\prime \xi_e(t^\prime)
   n_{DT}^2(t-t^\prime)
 \langle \sigma v(t-t^\prime) \rangle  \nonumber  \\ \vspace{+8.truemm}  &&  - 
\ A_b Z_{\hbox{\footnotesize eff}} n^2_e T_e^{1/2}
    +  \ \eta j^2
 - \ {3 \over 2} n_e ( T_e - T_i) / \tau_{\hbox{\footnotesize ei}} \ - \
  \nabla \cdot \vec \Gamma_{\hbox{\footnotesize E},e}
\end{eqnarray}

\vspace{-4.truemm}
\noindent and
                                                                                
\begin{eqnarray}
\hspace{-15truemm}  {\partial \over \partial t} \biggl[  {3 \over 2} ( n_{DT} &&  +   n_\alpha
  +  n_{\hbox{\footnotesize Be}} +
n_{\hbox{\footnotesize Ar}}  ) T_i \biggr]    =
\  P_{\hbox{\footnotesize aux},i} \ + \
{3 \over 2} n_e ( T_e - T_i) / \tau_{\hbox{\footnotesize ei}}
   \ - \ \nabla \cdot
\vec \Gamma_{\hbox{\footnotesize E},i} \ +
  \nonumber \\ \vspace{+6.truemm}
 & & \hspace{-4truemm}
{1 \over 4} f_{\hbox{\footnotesize  frac}}
\ f_i \ Q_\alpha
 \int_0^\infty
 dt^\prime \xi_i(t^\prime) n_{DT}^2(t-t^\prime)
 \langle \sigma v(t-t^\prime) \rangle
 \quad ;
\end{eqnarray}

 Here
$\vec \Gamma_{\hbox{\footnotesize DT}}$, $\vec \Gamma_\alpha$,
$\vec \Gamma_{\mbox{\footnotesize E},e}$ and
 $\vec \Gamma_{\mbox{\footnotesize E},i}$
are the DT and $\alpha$ particle fluxes and
the electron and ions energy fluxes due to transport, respectively.
 The  coefficients $A_b$, $\eta$ and
 $j$ correspond respectively,
 to the bremsstrahlung
radiation losses, the neoclassical resistivity and the toroidal
 plasma current density. $Z_{\hbox{\footnotesize eff}}$ is the effective
charge densiy; and $\tau_{\hbox{\footnotesize ei}}$ is
the relaxation time between the
energy densities of  the  electrons and the
ions.
The energy carried by the fusion alpha particles is $Q_\alpha = 3.5$ Mev;
$f_{\mbox{\footnotesize  frac}}$ is the effective fraction of alpha particles
not anomalously lost  during
thermalization;
$f_e$ and $f_i$ are the fraction  of the alpha particles  energy
$Q_\alpha$, deposited to the electrons and to the ions, respectively.
The thermalization of the alpha particles produced by fusion is not
assumed instantaneous but  time dependent with a distribution density
function
given by $\xi_{\hbox{\footnotesize th}}(t)$ for an alpha particle
produced at $t=0$. Similarly,
 the energy lost to the
electrons and the ions during the thermalization
process are also taken to be  time dependent
 following the distribution functions $\xi_e(t)$
and $\xi_i(t)$, respectively.  

The dynamical  equations
used in this work are the volume-averaged   of the above equations,
 assuming a
time dependent but homogeneous particle density throughout the plasma
with temperature radial profiles of the form[4]
                                                                                
\begin{equation}
T(\vec r, t) = T_0(t) [ 1 - (r/a)^2]^{\gamma_t} \, ,
\end{equation}
                                                                                
\noindent
 with $T_0$  the
 peak or central temperature, and $a$ the tokamak's minor radius.
 The radial profile parameter will be taken ${\gamma_t}=1.85$ for
 both the electrons
 and  the ions.
Transport losses are taken into account in the 0-D model
through the energy confinement time $\tau_E$, as well as by the D-T and
the helium ash confinement times $\tau_p$ and $\tau_\alpha$, respectively.

 The  nominal operating  state is assumed to be
 $n_0 = 1.01 \times
10^{20}$ m$^{-3}$ for the electron density;
 and $T_{\hbox{\footnotesize e}0}^{(n)} = 23.6 $ keV together with
$T_{\hbox{\footnotesize i}0}^{(n)} = 23.0 $ keV, for the central temperatures
of the electrons and the ions, respectively. The 
 helium ash fraction  nominal value is $f_0 = 0.045$. 
The relative fractions of the Be and Ar impurities are assumed i
$f_{\hbox{\footnotesize Be}}=0.02$ and $f_{\hbox{\footnotesize Ar}}= 0.0012$.
The ionization charge  will be assumed $Z_{\hbox{\footnotesize Ar}}
 = 17$, and $Z_{\hbox{\footnotesize Be}} = 4$.
The coefficient  $f_{\hbox{\footnotesize frac}}$ is assumed
 constant and equal to 0.9.
The above values of the plasma parameters 
 will constitute
 the  operating point for the
 ITER-FEAT like tokamak reactor used in this work.[5,6]
Here, we will assume that energy and particle scaling laws
are independent, but the DT and alpha particle confinement times
have a constant relationship, $ \tau_p = 0.6  \tau_\alpha$.

In practice, actual control actions are always constrained
between a maximum
and a minimum value, thus we shall impose  in the model described in Eqs.
(1)-(4) that
\begin{eqnarray}
0 \le  S_{f}^{\hbox{\footnotesize total}}  \le  2.3 \times 10^{22}
\ \hbox{sec}^{-1} ~, &
 \quad 0 \le  S_{\alpha}^{\hbox{\footnotesize total}}  \le  5.7 \times 10^{20}
 \ \hbox{sec}^{-1} ~, \nonumber \\ \vspace{+2.0truemm}
 \quad 0 \le  P_{\hbox{\footnotesize aux},e}^{\hbox{\footnotesize total}} 
 \le  95.2 \ \hbox{MW}
  \quad  & \hbox{and}
\quad  0 \le  P_{\hbox{\footnotesize aux},i}^{\hbox{\footnotesize total}} 
 \le 92.8 \ \hbox{MW}  
   ~~ ;
\end{eqnarray}
these limits  contain the required values for steady state
operation for the range of confinement times considered here. The plasma
core volume is assumed 837 m$^3$.

Assuming quasineutrality we have , $ n_e = n_{DT}
+  2 n_\alpha +
Z_{\hbox{\footnotesize Be}} n_{\hbox{\footnotesize Be}} +
Z_{\hbox{\footnotesize Ar}} n_{\hbox{\footnotesize Ar}}$;
and after taking
 volume average in Eqs. (1)-(4),
  we obtain a  coupled
set of nonlinear differential equations for the time dependence of
the electron density $n_e$, the  helium ash fraction
$ f_\alpha= n_\alpha/n_e $, and the peak electron and ions temperatures,
$T_{e0}$
and $T_{i0}$. Transport losses are taken into account in the resulting
 equations
through $\tau_E$, the energy confinement time, as well as by the DT and
helium ash confinement times $\tau_p$ and $\tau_\alpha$, respectively.

As pointed out in the Introduction, during the
thermalization
process, approximately 85\% of the energy of the fusion alphas
is absorbed by the electrons and only 15\% by the ions.
Thus, in this work we take $f_e=0.85$ and $f_i= 0.15$.
On the other hand,  for the nominal operating plasma parameters
 of the ITER-FEAT design, the time required by
 the alphas to reach the threshold energy
of 0.5 MeV, below which the energy is deposited mainly to the ions,
 for the nominal operating plasma parameters of the ITER-FEAT design
 is approximately 0.18 seconds; afterwards its energy is mainly deposited
to the ions, taking an additional 0.06 seconds approximately
to completely thermalize to the volume average plasma temperature of
approximately 8.0 keV. 

In order to stabilize the system around a given state, the neural
network must provide
appropriate values for the control variables, according to the
current state of the system.
In all the simulated transients used in the training and testing of the
 neural network in this and the next sections,
 we use
a fourth order Adams-Moulton integration scheme with two corrector-predictor
steps, using a constant time step of length  0.02 sec.
 The control actions are updated every 0.06 sec; in other words,
 the values of the control
variables in Eqs. (6)  remain the constant for three consecutive
time steps and then updated, feeding to the RBNN  the current values
of the electron density,
the fraction of helium ash, the ion and electron peak temperatures and
the energy and helium ash confinement times.

\section{Simulation Results}

We present here an example  of a  typical transient
behaviour
 with  the resulting  network controller, obtained after training
 the RBNN using a backpropagation through time algorithm.[7]
The tokamak reactor is assumed to follow 
 the  IPB98(y,2) scaling law,[8] i.e.
                                                                                
\begin{equation}
\tau_{\hbox{\footnotesize IPB98}} = 0.056 I^{0.93} R^{1.97} B^{0.15} M^{0.19}
\epsilon^{0.58} \kappa^{0.78} n_e^{0.41} P^{-0.69}_{net}  \quad ;
\end{equation}

\noindent and the   ratio $r= \tau_\alpha/ \tau_E $ will be
 assumed to randomly fluctuate following a Gaussian distribution with mean 
value value $\bar r=4.5$ with standard deviation $0.04\times \bar r$;   
while its "on-line" estimation will also be a Gaussian stochastic variable
 with
the same mean value but with standard deviation $0.08\times \bar r$. 
In the  transient shown below we choosed the following initial
conditions $n_e = 1.15 \times n_0$ for the electron density;
 $f_\alpha = 0.80 \times f_0$, which
corresponds to a helium ash density of 8 \% below its nominal value;
and an initial peak electron  and ion temperatures of
 $T_e = 1.15 \times T_{0e}^{(n)}$ and
$T_i = 1.15 \times T_{0i}^{(n)}$, respectively.
In Figures 1 and 2 we show the behaviour of the normalized  electron
density, helium ash fraction, the electron temperature and the
ions temperature, as function into the transient. In Figures 3 and 4 
we show the time behaviour of the control variables, normalized with
respect their maxima allowable values, as function ito the transient.
It is observed that the RBNN controller is able to supress these
fluctuations within 12 seconds into the transient.
In Figure 5 (left) we show the time behaviour of the IPB98(y,2), Eq. (7),
for this transient; and in Fig. 5 (right) the random fluctuations
of the "on-line" estimation of the ratio $\tau_\alpha/\tau_E$, along
the duration of the transient.

%------------------------------------
\begin{figure}[ht]
\begin{minipage}[t]{7.8truecm}
\begin{center}
\mbox{\epsfig{file=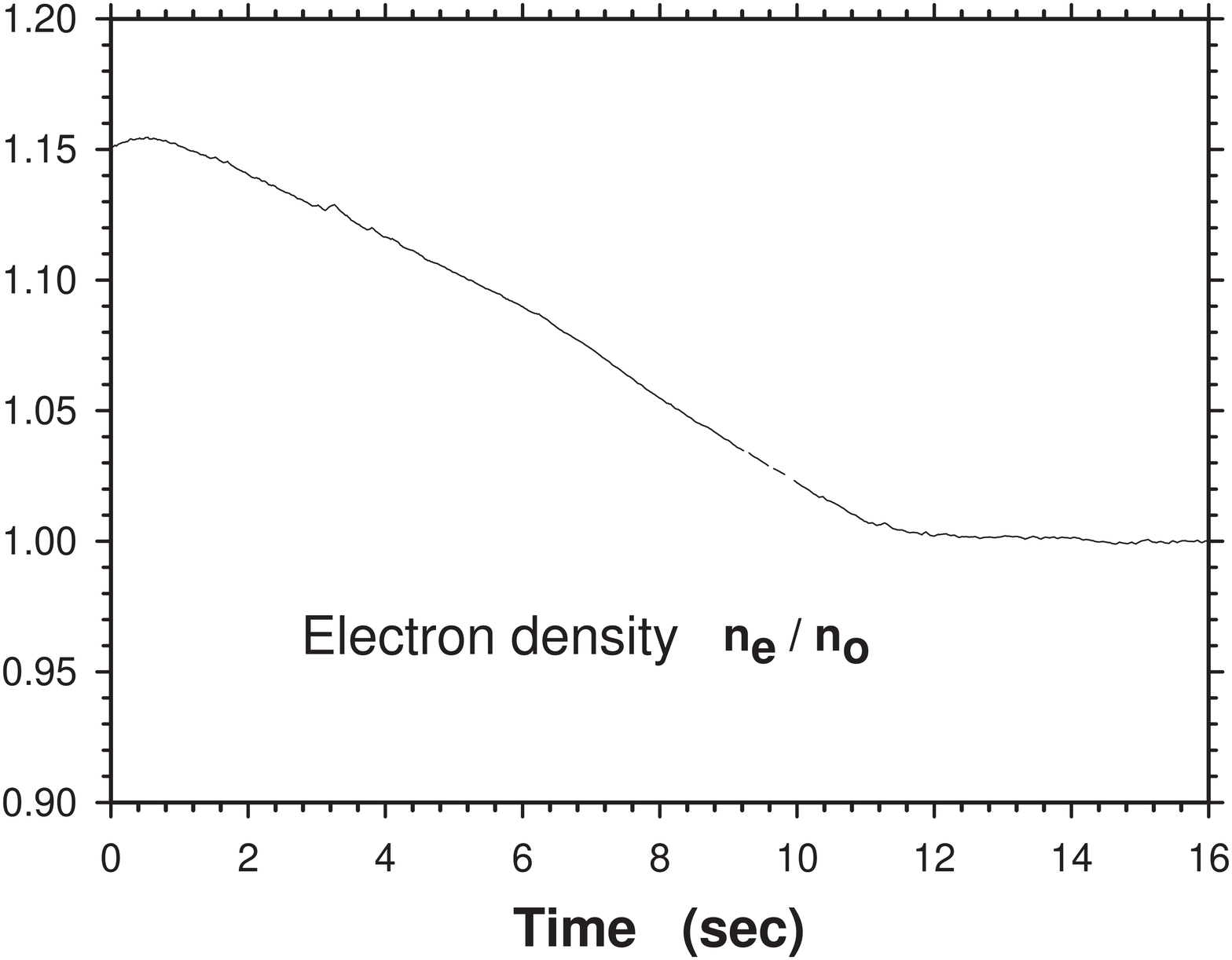,width=7.8truecm}}
\end{center}
\end{minipage}
\hfill
\begin{minipage}[t]{7.8truecm}
\begin{center}
\mbox{\epsfig{file=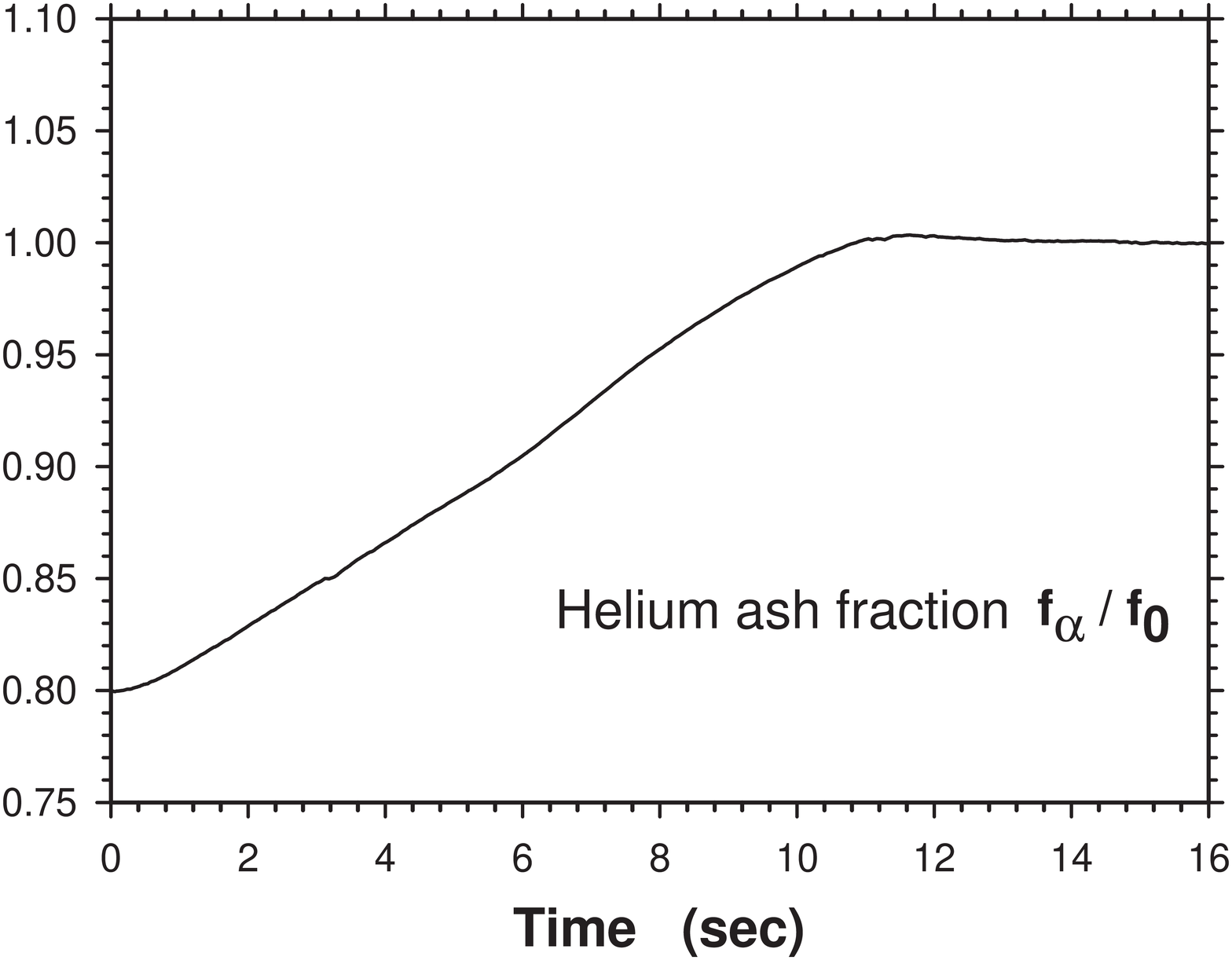,width=7.8truecm}}
\end{center}
\end{minipage}
\vspace{-10truemm}
\caption{ Behaviour of the electron density (left) and the
helium ash fraction
(right)  as function of time
corresponding to the transient described in the text. }
\end{figure}

%------------------------------------
\begin{figure}[h]
\begin{minipage}[t]{7.8truecm}
\begin{center}
\mbox{\epsfig{file=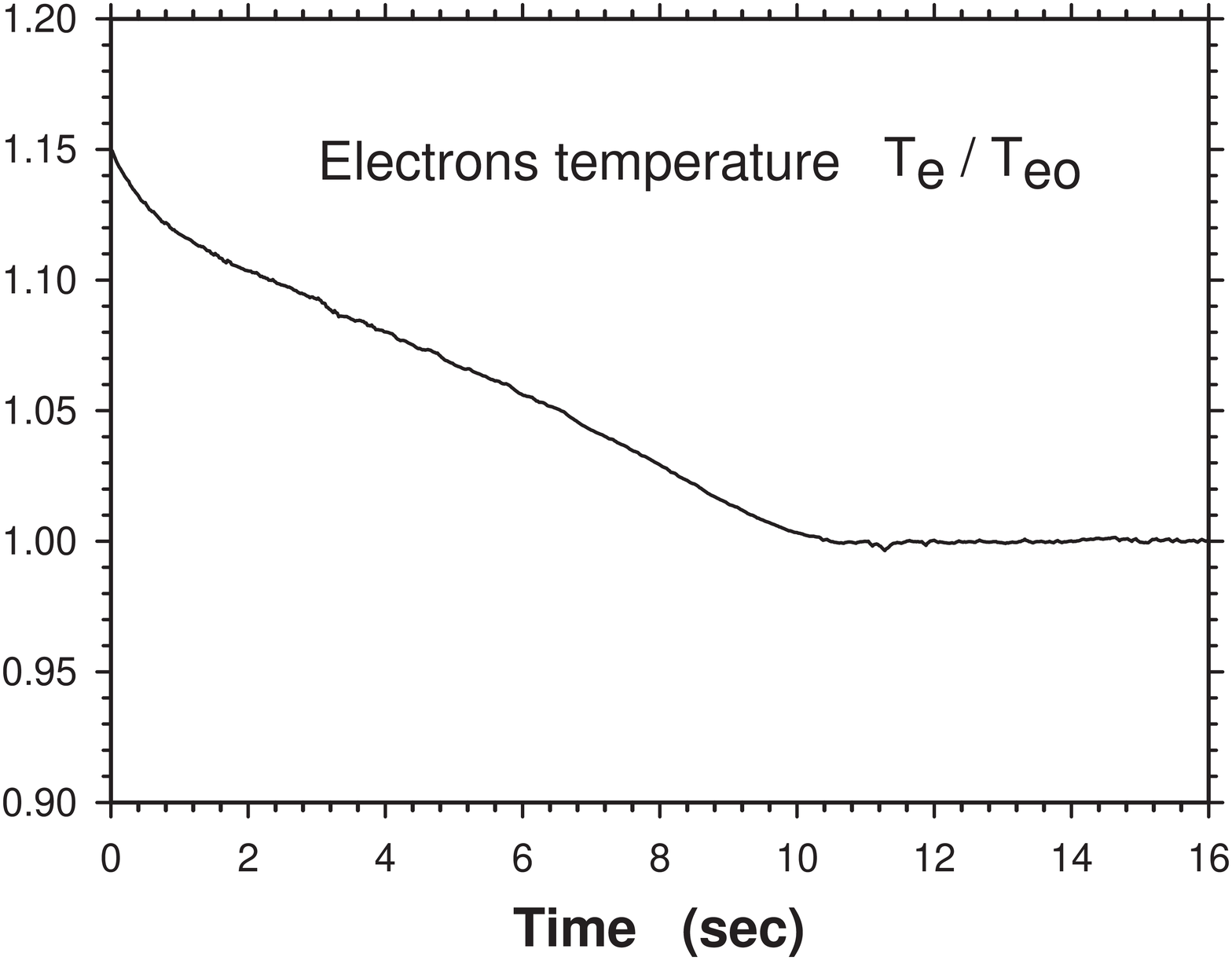,width=7.8truecm}}
\end{center}
\end{minipage}
\hfill
\begin{minipage}[t]{7.8truecm}
\begin{center}
\mbox{\epsfig{file=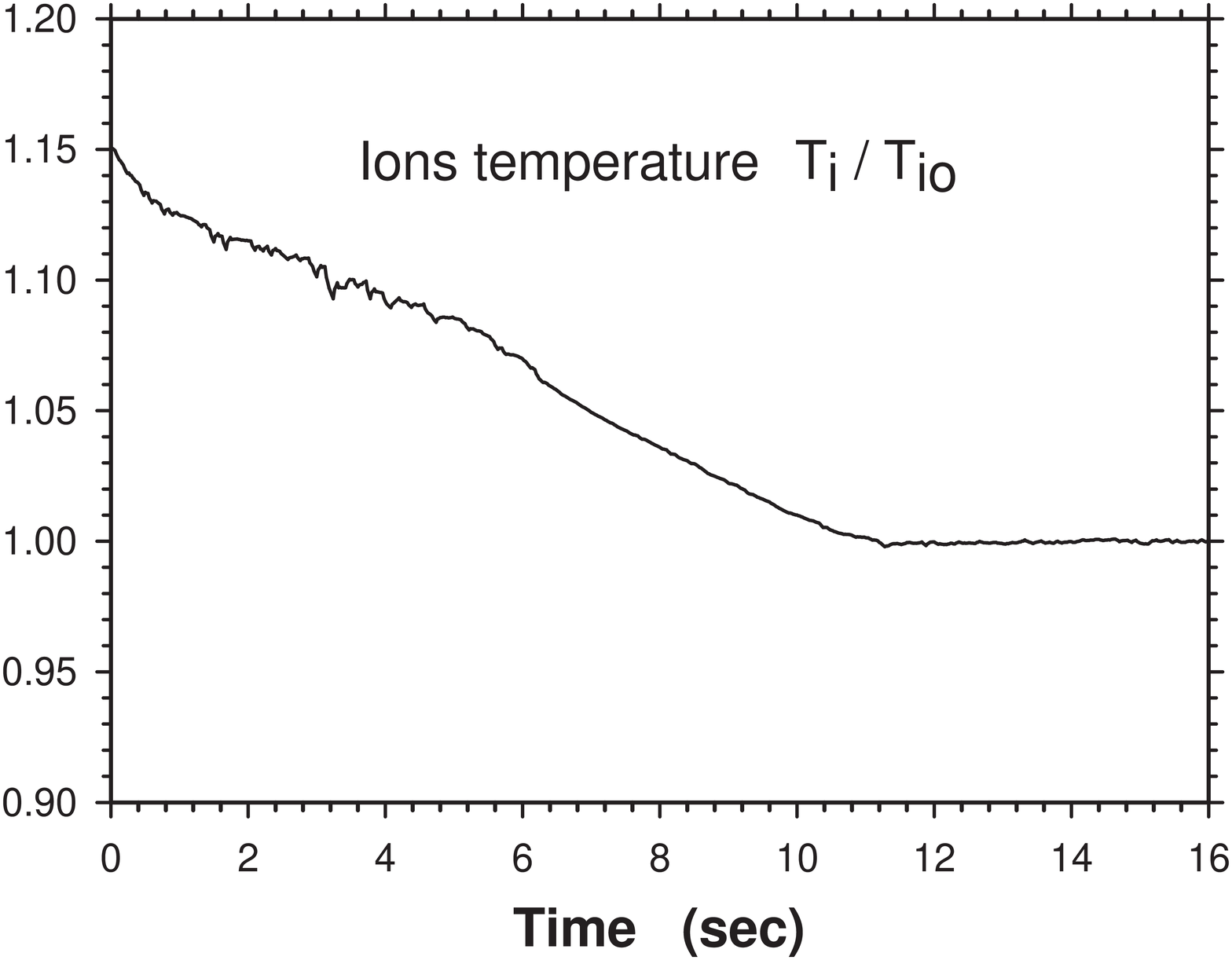,width=7.8truecm}}
\end{center}
\end{minipage}
\vspace{-10truemm}
\caption{  Behaviour of the electron 
and ions temperatures, left and right respectively, 
 as function of time
corresponding to the transient described in the text. }
\end{figure}
%------------------------------------

\begin{figure}[h]
\vspace{4truemm}
\begin{minipage}[t]{7.8truecm}
\begin{center}
\mbox{\epsfig{file=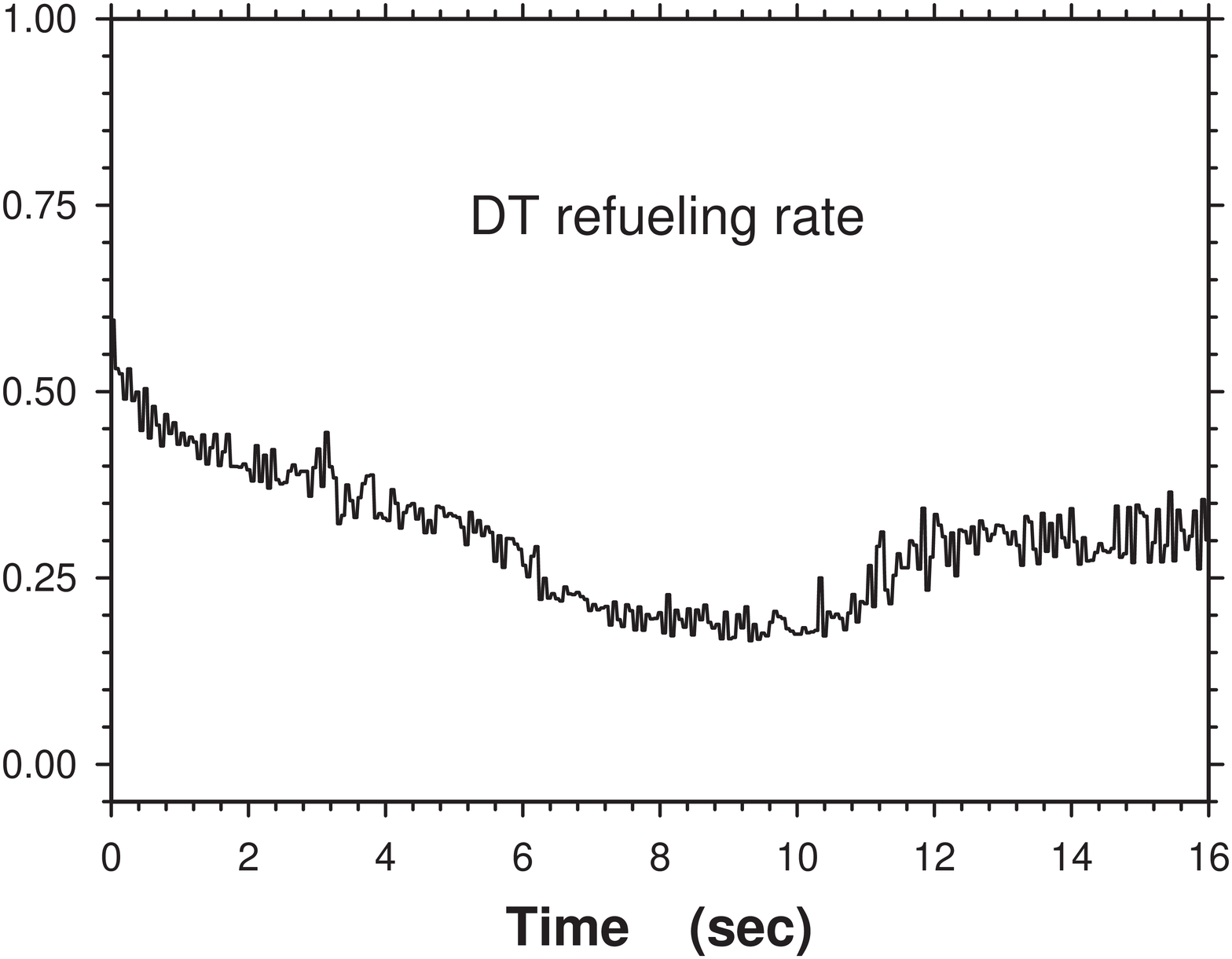,width=7.8truecm}}
\end{center}
\end{minipage}
\hfill
\begin{minipage}[t]{7.8truecm}
\begin{center}
\mbox{\epsfig{file=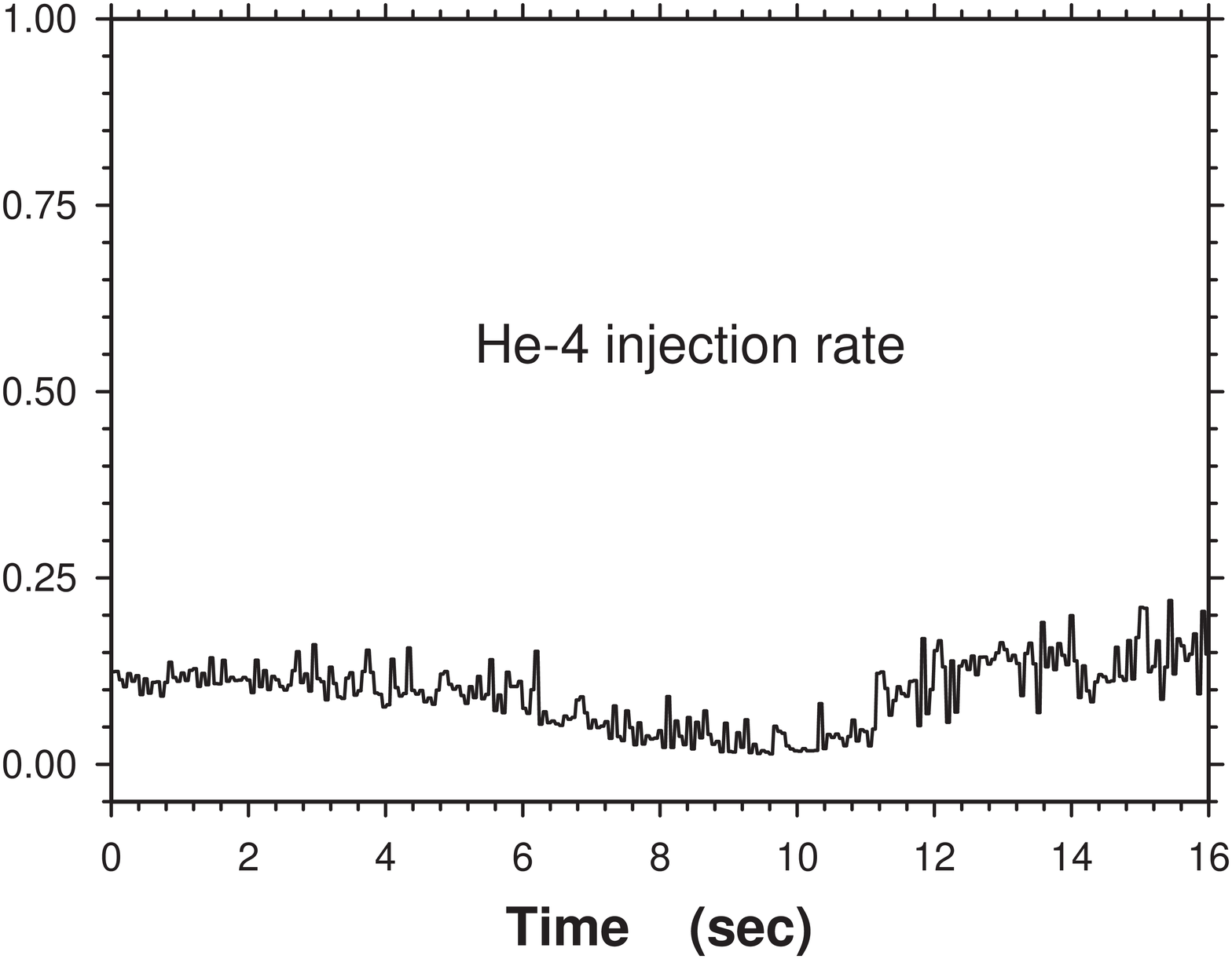,width=7.8truecm}}
\end{center}
\end{minipage}
\vspace{-10truemm}
\caption{  Normalized behaviour of the DT refueling rate (left) and 
neutral He-4 injection rate
(right) as function of time
corresponding to the transient described in the text. }
\end{figure}
%------------------------------------

\section{Conclusions}
We have shown
 that burn control of an ITER-FEAT like tokamak with uncertainties
in the helium
ash confinement time
can be succesfully achieved with radial basis neural networks.
Assuming the reactor follows  IPB98(y,2) scaling law,
and using a 0-D two temperature volume-averaged model we illustrate
by means of a typical transient that the RBNN controller is robust 
with respect to
  noisy "on-line"
 measurements of the
ratio $\tau_\alpha/\tau_E$.
A complete report of these results including "on-line" measurement
noise in the estimation of the energy confinement time is under
preparation.[9]

%------------------------------------
\begin{figure}[t]
\vspace{-3truemm}
\begin{minipage}[t]{7.8truecm}
\begin{center}
\mbox{\epsfig{file=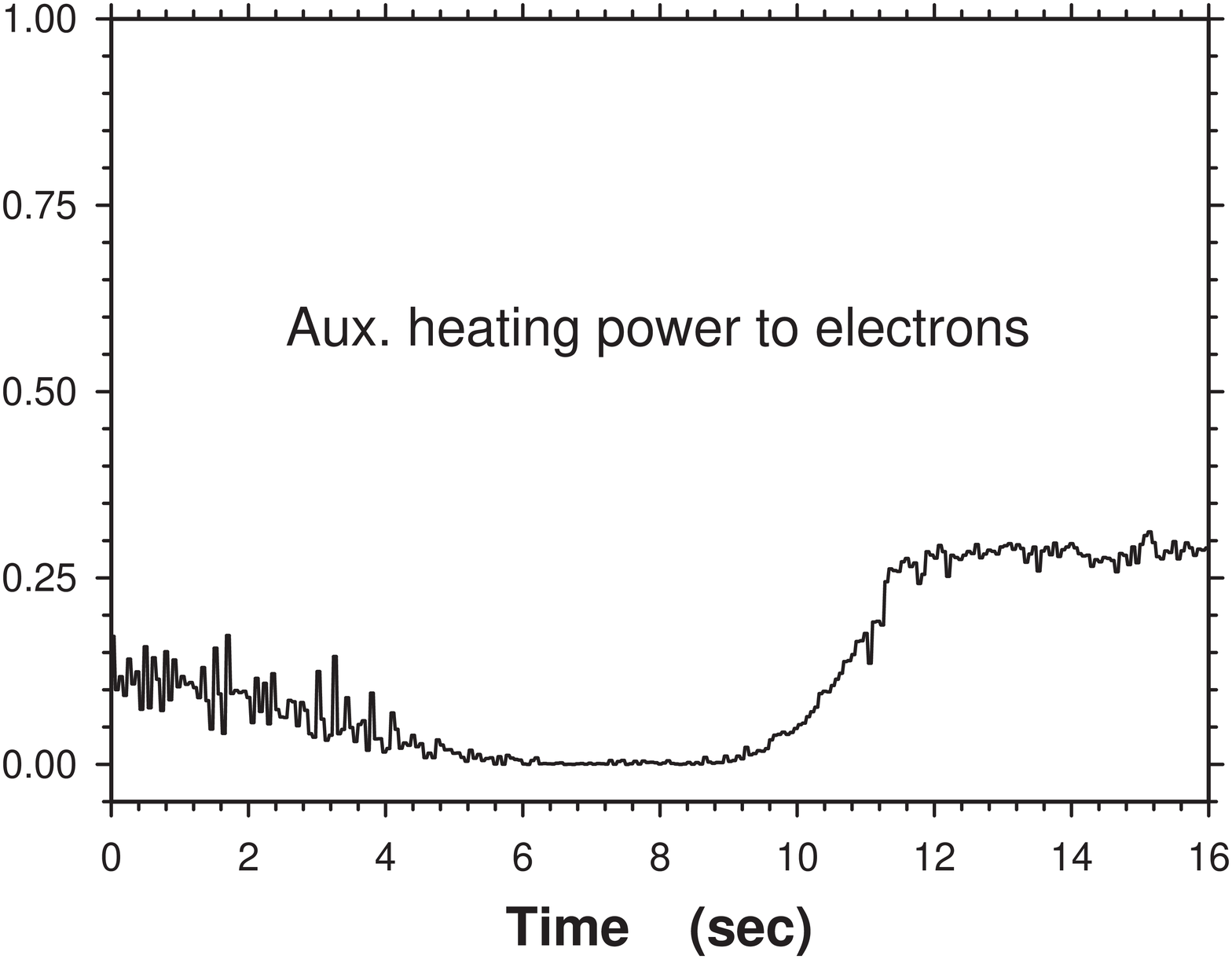,width=7.8truecm}}
\end{center}
\end{minipage}
\hfill
\begin{minipage}[t]{7.8truecm}
\begin{center}
\mbox{\epsfig{file=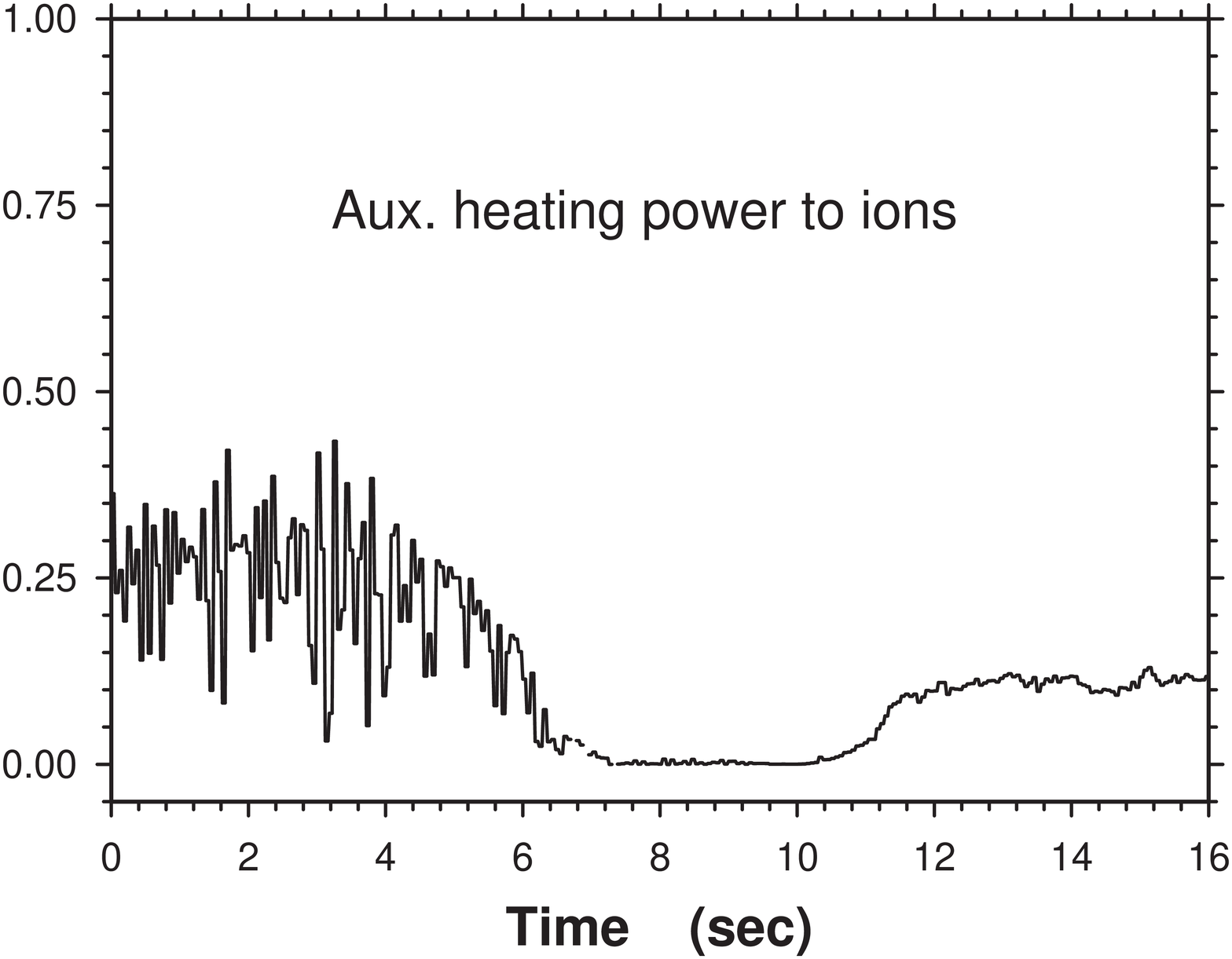,width=7.8truecm}}
\end{center}
\end{minipage}
\vspace{-10truemm}
\caption{ Normalized behaviour of the auxiliary heating power to electrons
 (left) and 
to ions 
(right) as function of time
corresponding to the transient described in the text. }
\end{figure}
%==============================================
\begin{figure}[t]
\vspace{-3truemm}
\begin{minipage}[t]{7.8truecm}
\begin{center}
\mbox{\epsfig{file=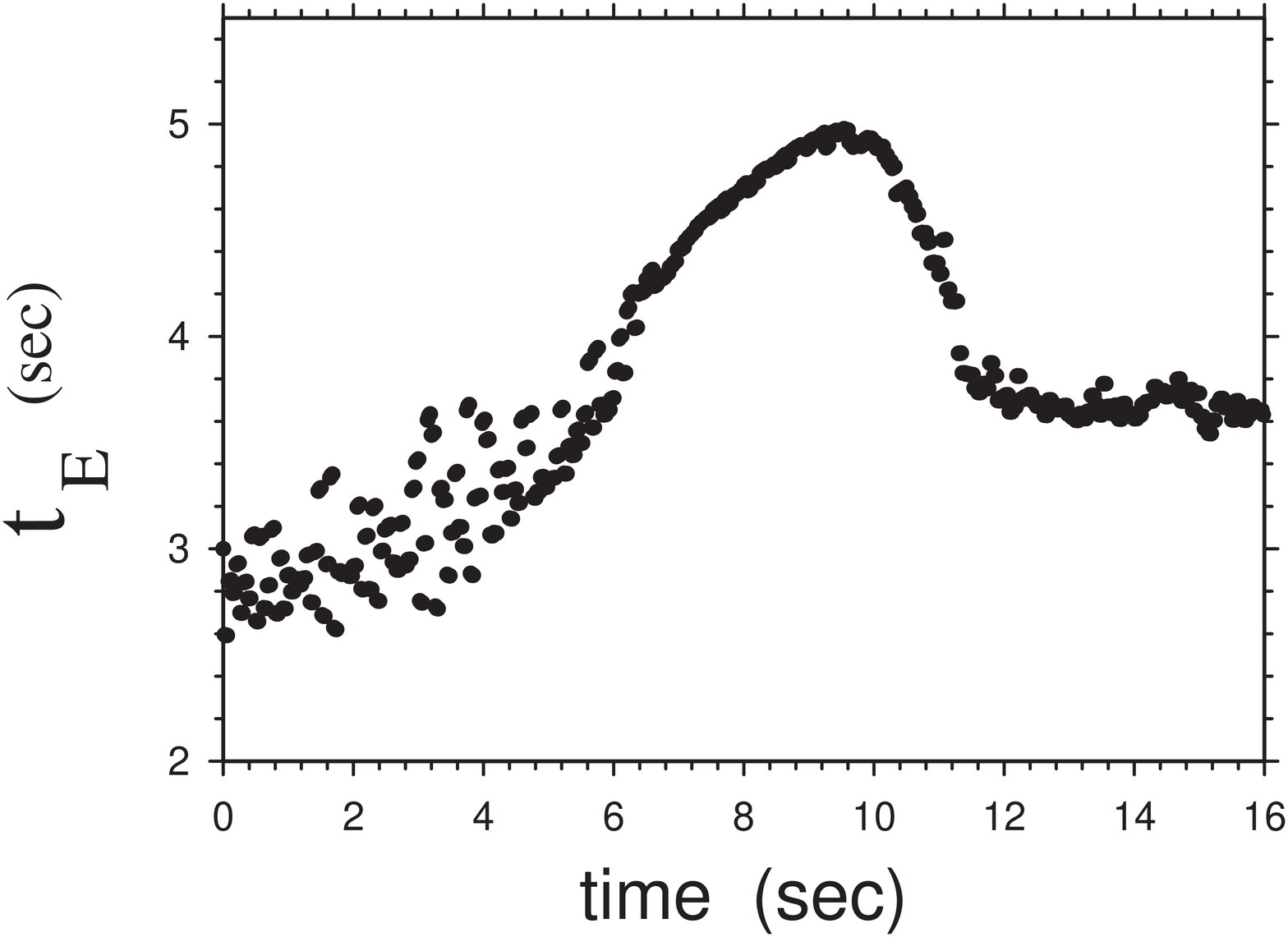,width=7.8truecm}}
\end{center}
\end{minipage}
\hfill
\begin{minipage}[t]{7.8truecm}
\begin{center}
\mbox{\epsfig{file=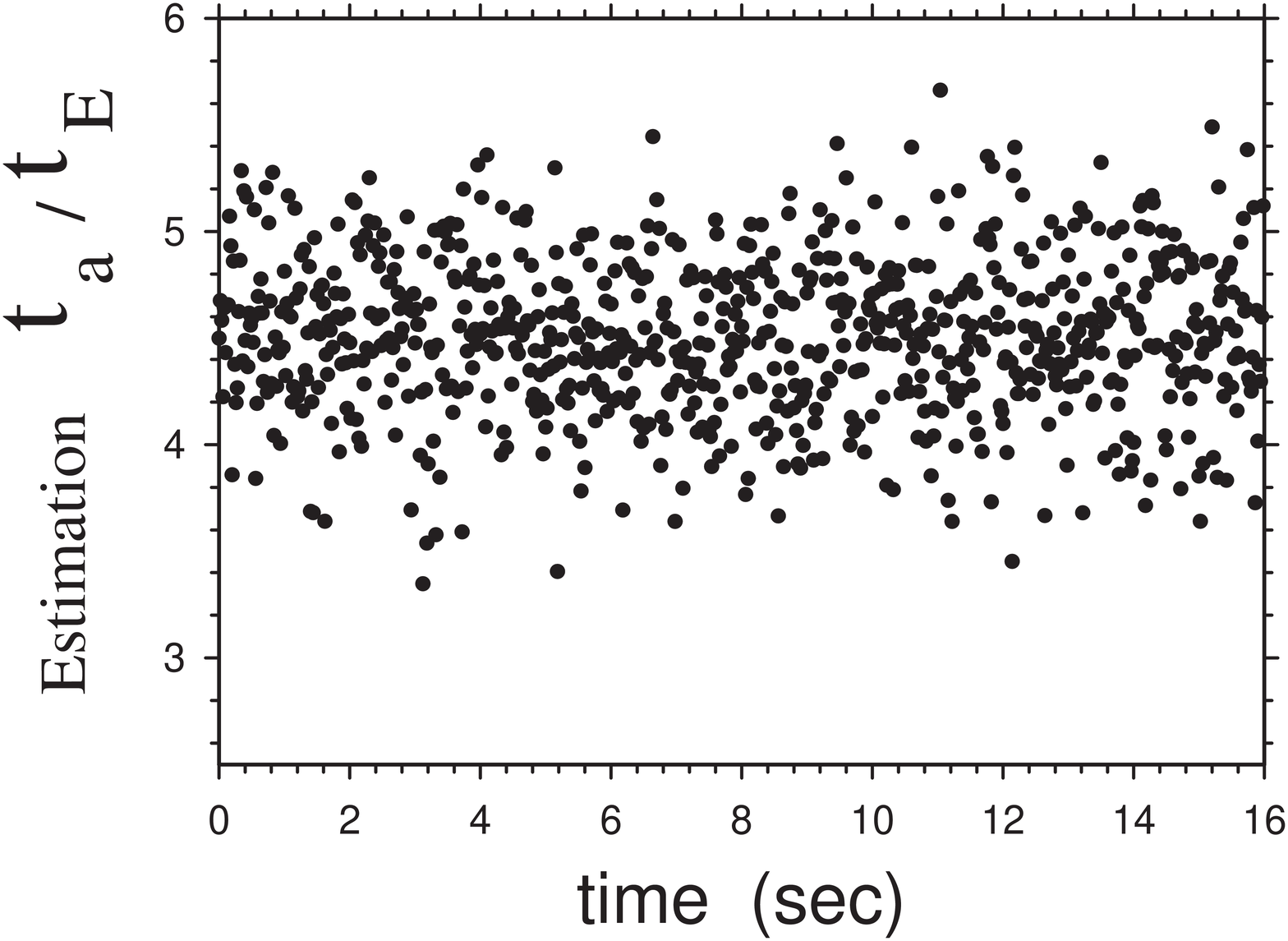,width=7.8truecm}}
\end{center}
\end{minipage}
\vspace{-10truemm}
\caption{
 Energy confinement time $\tau_E$ as obtained from Eq. (7) for the
IPB98(y,2) scaling  (left)
and the noisy "on-line" estimation of the ratio $\tau_\alpha / \tau_E$,
(right) used by the
RBNN to update the control variables for the transient discussed in the 
text.
}
\end{figure}
%===============================================

\

\

\vspace{+3truemm}

\section*{ Acknowledgments}
\vspace{-1truemm}
 Partial  financial  support from DGAPA-UNAM IN118505 project is
  gratefully acknowledge. The author also wishes to  thank the Department of
 Supercomputing at UNAM for allowing him access to the  multiprocessor
 AlphaServer SC 45.

\end{document}